\documentclass[twocolumn]{aastex631}
\pdfoutput=1
\usepackage{natbib}
\usepackage{graphicx}
\usepackage{hyperref}
\usepackage{float}

\def\be{\begin{equation}}
\def\ee{\end{equation}}
\def\bfnabla{\mbox{\boldmath $\nabla$}}

\begin{document}

\newcommand{\Cornell}{Department of Physics, Cornell University, Ithaca, NY 14853, USA}

\title{Khronometric theories of modified Newtonian dynamics}

\author{\'Eanna.~\'E.~Flanagan}
\affiliation{\Cornell}

\begin{abstract}

In 2011 Blanchet and Marsat suggested a fully relativistic version
of Milgrom's modified Newtonian dynamics (MOND) in which the dynamical
degrees of freedom consist of the spacetime metric and a foliation of
spacetime, the khronon field.
This theory is simpler than the alternative relativistic formulations.
We show that the theory has a consistent non-relativistic or slow motion limit.
Blanchet and Marsat showed that in the slow motion limit the theory reproduces stationary
solutions of modified Newtonian dynamics.
We show that these
solutions are stable to khronon perturbations in the low acceleration
regime, for the cases of spherical, cylindrical and planar symmetry.  
For non-stationary systems in the low acceleration regime we show that
the khronon field generally gives an order unity correction to the
modified Newtonian dynamics.

\end{abstract}

\section{Introduction}\label{sec:introduction}

In 1983 Milgrom 
suggested a modification of Newtonian gravity in order to provide a
better fit to Galactic rotation curves
\citep{1983ApJ...270..365M,1983ApJ...270..371M,1983ApJ...270..384M},
now known as Modified Newtonian Dynamics (MOND).
On galactic scales this theory fits the data well and provides an
explanation for a number of otherwise mysterious regularities \citep{Famaey:2011kh,McGaugh:2014nsa,Dainotti:2023yrk}.
Conventional Newtonian dynamics within the $\Lambda$CDM framework
has difficulty explaining these regularities and has other
difficulties when confronted with data on galactic structure \citep{Peebles:2010di}.

On the other hand MOND is clearly inconsistent with the data on
cluster scales, and in particular with observations of the bullet
cluster which exhibits a spatial separation between the dark and baryonic
matter in the aftermath of a collision \citep{2006ApJ...648L.109C}.   
The MOND theory requires the assumption of neutrino dark matter on
these scales to be consistent with these observations
\citep{Famaey:2011kh,McGaugh:2014nsa,Dainotti:2023yrk}.  These
neutrinos need not contribute appreciably on galactic scales due to
the Tremaine-Gunn bound \citep{1979PhRvL..42..407T}, {\it i.e.} Fermi degeneracy pressure.

The original formulation of MOND was valid only in the
non-relativistic limit, and it is necessary to have a fully
relativistic formulation for a number of reasons.  The first such
theory was suggested by \citet{2004PhRvD..70h3509B}, but this theory
has since been ruled out by LIGO observations which show that
gravitational and electromagnetic waves propagate at very nearly the
same speed \citep{2018PhRvD..97d1501B}.  A more recent theory that is compatible with LIGO data
is that of \citet{Skordis:2020eui,Skordis:2021mry}.  However this
theory is quite complicated with the dynamical fields consisting of a metric, a scalar and a unit-norm vector field.
In 2011 Blanchet and Marsat (henceforth BM) suggested a fully relativistic version
of MOND in which the dynamical
degrees of freedom consist of the spacetime metric and a foliation of
spacetime, the khronon field \citep{Blanchet:2011wv,Blanchet:2012ub,Sanders:2018jiv}.
Their theory is a special case of a general class of theories 
called khronometric theories which includes Horava gravity
\citep{Horava:2009uw,Blas:2009qj,Blas:2014aca,Blas:2010hb}.
They can be obtained as a limiting case 
of a class of theories with a dynamical unit timelike vector field,
which includes the Einstein-Aether theory \citep{Jacobson:2010mx,Jacobson:2013xta}.

In this paper we show that the BM theory has a consistent
non-relativistic or slow motion limit, 
of a kind different to that previously considered in the literature
in the context of a more general class of theories in which the limit
to the BM theory is pathological \citep{Bonetti:2015oda}.

We also study the dynamics of the khronon field in the BM theory.
\citet{Blanchet:2011wv} showed that in the slow motion limit the theory reproduces stationary
solutions of modified Newtonian dynamics with the khronon perturbation
set to zero in a certain coordinate system.  We show that these
solutions are stable to khronon perturbations in the low acceleration
regime, for the cases of spherical, cylindrical and planar symmetry.  
We also show that 
for non-stationary systems in the low acceleration regime 
the khronon field generally gives an order unity correction to the
modified Newtonian dynamics.

\vspace{0.5cm}

\section{The khronon theory of Blanchet and Marsat}
\label{sec:BMtheory}

In khronometric theories of gravity 
the dynamical fields are a metric $g_{ab}$ and a spacelike spacetime foliation.
The foliation is encoded in a scalar field $T$, called the khronon field, and the theory is
invariant under $T \to h(T)$ where $h$ is any monotonic function.
The unit, future directed vector field ${\vec n}$ normal to the foliation is given
\be
n_a = \frac{- \nabla_a T} {\sqrt{- g^{ab} \nabla_a T \nabla_b T}},
\label{ndef}
\ee
and we can decompose its derivative as
\be
\nabla_a n_b = - n_a a_b + \frac{1}{3} \theta (g_{ab} + n_a n_b) + \sigma_{ab},
\label{adef}
\ee
where $a^a = n^b \nabla_b n^a$ is the acceleration, $\theta = \nabla_a
n^a$ is the expansion and $\sigma_{ab}$ satisfying $n^a \sigma_{ab} =
0$, $\sigma_{[ab]}=0$ is the shear.
The action of the theory is a
function of $a = \sqrt{a^a a_a}$, $\theta$ and $\sigma = \sqrt{\sigma_{ab} \sigma^{ab}}$, all of which have dimensions of inverse length.

We focus on the particular theory suggested by 
\citet{Blanchet:2011wv} and \citet{Blanchet:2012ub}
as a fully relativistic version of MOND.
The action for the theory is
\be
S = \frac{c^3}{16 \pi G} \int d^4 x \sqrt{-g} \left[ R - 2 f(a) \right]
+ S_{\rm m}[g_{ab}, \Psi]
\label{action0}
\ee
where $S_{\rm m}$ is the matter action, $G$ is Newtons constant and
$c$ the speed of light.  Here the function $f(a)$ 
satisfies
\be
f(a) = \Lambda_0 - a^2 + \frac{2 c^2 a^3}{3 a_0} + O(a^4), \ \ \ \ a
\ll a_0
\label{fdef}
\ee
where $a_0 = 1.2 \times 10^{-2} {\rm m} {\rm s}^{-2}$ is the MOND
acceleration scale,
and
\be
f(a) \to \Lambda_\infty, \ \ \ \ a \gg a_0.
\label{fdef1}
\ee
for some constants $\Lambda_0$ and $\Lambda_\infty$.
The equations of motion for this theory are \citep{Blanchet:2011wv}
  \be
  G^{ab} =  \frac{8 \pi G}{c^4} \left[ T_{\rm m}^{ab} + T_{T}^{ab}
    \right],
  \label{ee}
  \ee
  where $T_{\rm m}^{ab}$ is the matter stress energy tensor, and
  \be
  T_T^{ab} = - \frac{c^4}{8 \pi G} \left[ f g^{ab} + 2 n^a n^b \nabla_c (\chi a^c) - 2 \chi a^a a^b \right]
\label{khrononTab}
  \ee
  is the stress energy tensor of the khronon field, with
  \be
  \chi(a) = \frac{f'(a) }{2 a}.
  \label{chidef}
  \ee
  The equation of motion for the khronon field is
  \be
  \nabla_a \left[ n^a \nabla_b(\chi a^b)\right] = \frac{1}{2} n^a
  \nabla_a f + \chi a^a a^b \nabla_a n_b.
  \label{eom2}
  \ee

\section{Slow motion limit}
\label{sec:slomo}

We now specialize to the slow motion limit, which we define as follows.
Consider an isolated source of gravity characterized by a mass scale
${\cal M}$, a lengthscale ${\cal L}$ and a timescale ${\cal T}$. We
define the dimensionless quantities
\be
\label{hatcdef}
   {\hat G} = \frac{G {\cal M} {\cal T}^2 }{ {\cal L}^3}, \ \ \ \ 
   {\hat c} = \frac{c {\cal T}}{{\cal L}}, \ \ \ \
   {\hat a}_0 = \frac{ a_0 {\cal T}^2}{{\cal L}}.
\ee
The slow motion limit is the limit ${\hat c} \to \infty$ at fixed ${\hat
  G}$ and fixed ${\hat a}_0$.
In particular the slow  motion limit encompasses both the regime
${\hat  a}_0 \ll 1$ where normal Newtonian gravity is recovered, and the
regime ${\hat a}_0 \gg 1$ where modified Newtonian dynamics
applies\footnote{The timescale ${\cal T}$ is of course not independent
of the other parameters when the source is self-gravitating.  In the
Newtonian regime ${\hat a}_0\ll 1$ this occurs when ${\hat G} \sim 1$
or ${\cal T} \sim G^{-1/2} {\cal L}^{3/2} {\cal M}^{-1/2}$.  In the
MOND regime ${\hat a}_0 \gg 1$ the condition is instead ${\hat G}
{\hat a}_0 \sim 1$ or ${\cal T} \sim G^{-1/4} a_0^{-1/4} {\cal L}
{\cal M}^{-1/4}$.\label{calT}}.  Note that this limit is usually
called the Newtonian limit in the context of general relativity; we
will generally avoid that terminology here since conventional
Newtonian dynamics are not recovered in the limit ${\hat c} \to
\infty$ unless ${\hat a}_0 \ll 1$.

To examine the slow motion limit we specialize to conformally
Cartesian harmonic gauge, with a metric expansion of the standard form 
\begin{eqnarray}
  ds^2 &=& - c^2 \left[ 1 + \frac{2 \Phi}{c^2} +
 \frac{2 \Phi^2}{c^4} + \frac{2 \psi}{c^4} 
+    O(c^{-6}) \right] dt^2
\nonumber \\
&&+ \frac{2}{c^2}\left[ \zeta_i + O(c^{-2}) \right] dx^i dt \nonumber \\
&& + \left[ \delta_{ij} - \frac{2 \Psi}{c^2} \delta_{ij} + O(c^{-4})
  \right] dx^i dx^j,
\label{metricPN}
\end{eqnarray}
with the leading order potential $\Phi$ and subleading potentials
$\Psi$, $\psi$ and $\zeta_i$.
We assume an expansion of the khronon field of the form
\be
T(t,x^i) = t + \frac{1}{c^2} \pi(t,x^i) + O(c^{-4}).
\label{Texpand}
\ee
where the khronon perturbation $\pi$ has dimensions of
$[L]^2[T]^{-1}$.  The motivation for the ansatz (\ref{Texpand}) is
that it guarantees that the khronon stress energy tensor
(\ref{khrononTab}) has scalings with $c$ of the standard post-Newtonian
form (see the appendix)
\be
\label{standard}
T^{tt} = O(c^0), \ \ \ T^{ti} = O(c^0), \ \ \ T^{ij} = O(c^0),
\ee
consistent with the metric ansatz (\ref{metricPN}).
This in turn, when combined with the Einstein equation (\ref{ee}),
implies the equality of the potentials $\Phi$ and $\Psi$ in the metric
(\ref{metricPN}):
\be
\Psi = \Phi.
\label{equality}
\ee

To explore the dynamics in the slow motion limit it is easiest to work
with the action (\ref{action0}) rather than the equations of motion
(\ref{ee}) and (\ref{eom2}).
It is convenient to define a rescaled acceleration variable
\be
   {\bar a} = c^2 a
   \label{rescale}
   \ee
which has units of acceleration, and a rescaled function of
acceleration
\be
   {\bar f}({\bar a}) = c^4 f({\bar a}/c^2).
   \label{barfdef}
\ee
By inserting the ansatzes (\ref{metricPN}) and (\ref{Texpand}) and the result (\ref{equality}) into the definitions
(\ref{ndef}), (\ref{adef}) and (\ref{rescale}) we
obtain the expansions (see the appendix)
\begin{eqnarray}
   {\bf n} &=& - c \left[ 1 + \frac{\Phi}{c^2} + \frac{ ({\bfnabla} \pi)^2 }{2 c^2} + O(c^{-4}) \right] dt \nonumber \\
   &&   - \left[ \frac{\pi_{,i}}{c} + O(c^{-3}) \right] dx^i,
   \label{n1}
\end{eqnarray}
and
\be
{\bar a}^2 =  ({\bfnabla \Xi})^2 + O(c^{-2}),
\label{a2}
\ee
where
\be
\Xi = \Phi - {\dot \pi} + \frac{1}{2} ( {\bfnabla} \pi)^2
\label{Xidef}
\ee
and where ${\bfnabla}$ is the spatial gradient.  The nonrelativistic
limit of the action (\ref{action0}) becomes\footnote{The general form of the dependence of this action on $\pi$ is dictated by Galilean invariance, and the same class of actions is obtained for a superfluid coupled to Newtonian gravity \citep{2021arXiv210910928K}.}
\be
S = -\int dt \int d^3 x \left[   \frac{(\bfnabla \Phi)^2}{8 \pi G}
+  \rho_{\rm m} \Phi + \frac{1}{8 \pi G} {\bar f}({\bar a})  \right] + O(c^{-2}),
\label{action1}
\ee
where $\rho_{\rm m}$ is the mass density.

Varying
with respect to the khronon perturbation
$\pi$ gives the khronon equation of motion in the slow motion limit in the form of a continuity equation
\be
   {\dot \rho}_T + \bfnabla \cdot ( \rho_T {\bf v}_T) =0.
   \label{khrononeom}
   \ee
Here the khronon mass density is
   \be
   \rho_T = - \frac{1}{4 \pi G} \bfnabla \cdot \left[ {\bar \chi}({\bar a})
     \bfnabla \Xi\right]
   \label{rhoTdef}
   \ee
where
\be
{\bar \chi}({\bar a}) = \frac{ {\bar f}'({\bar a}) }{2 {\bar a}} =
\chi({\bar a}/c^2),
\label{barchidef}
\ee
and its velocity is
   \be
      {\bf v}_T = - {\bfnabla} \pi.
      \label{vTdef}
      \ee

Varying  the action (\ref{action1}) with respect to the Newtonian potential $\Phi$ and using
Eqs.\ (\ref{chidef}), (\ref{a2}) and (\ref{Xidef}) gives
\be
\nabla^2 \Phi + \bfnabla \cdot \left[ {\bar \chi}({\bar a}) \bfnabla \Xi\right] = 4 \pi G
\rho_{\rm m}.
\label{eom0}
\ee
This reduces to the MOND equation for $\Phi$ when $\pi=0$, as shown by BM:
\be
\bfnabla \cdot \left[ (1 + {\bar \chi}({\bar a}_{\rm b})) \bfnabla \Phi\right] = 4 \pi G
\rho_{\rm m},
\label{eom1}
\ee
where we have defined the background acceleration
\be
   {\bar a}^i_{\rm b} = \partial_i \Phi, \ \ \ \
   {\bar a}_{\rm b} = | \bfnabla \Phi|.
   \ee

\subsection{Unitary gauge}
\label{sec:unitary}

Starting from the conformally Cartesian
harmonic gauge used here, 
it is possible to set the khronon perturbation to zero by a change of gauge
of the form
\be
t \to {\bar t} = t + \pi/c^2,
\ee
yielding from (\ref{Texpand}) that
\be
T({\bar t},x^i) = {\bar t} + O(c^{-4}).
\label{Texpand1}
\ee
In this so-called unitary gauge the form (\ref{metricPN}) of the metric expansion is altered,
and in particular the ${\bar t}i$ component of the metric has a term at $O(c^0)$,
\be
g_{{\bar t}i} = \pi_{,i} + O(c^{-2}),
\label{gmexpand}
  \ee
  rather than being of order $O(c^{-2})$ as is normally the case.

It should be possible to rederive all the results of this paper using
the unitary gauge (\ref{Texpand1}).  However, it is somewhat awkward to
use this gauge, since it is incompatible with the standard post-Newtonian
expansion framework and so requires generalizing this framework.
This awkwardness is 
why we chose
conformally Cartesian
harmonic gauge.  However we emphasize that there is no issue of principle with 
the use of unitary gauge.

\subsection{Dynamics of the khronon field}

Some insight into the dynamics of the khronon field can be obtained by
linearizing the equation of motion (\ref{khrononeom}) in $\pi$.
We obtain
\be
\partial_i ( h^{ij} \partial_j {\ddot \pi}) + \partial_i (\rho_{T0}
\partial_i \pi) = {\dot \rho}_{T0} + O(\pi^2).
\label{dynamics}
\ee
Here the tensor $h^{ij}$ is given by
\be
h^{ij} = - \frac{1}{4 \pi G} \left[ {\bar \chi}({\bar a}_{\rm b}) (\delta^{ij}
  - {\hat a}_{\rm b}^i {\hat a}_{\rm b}^j) + {\bar f}''({\bar a}_{\rm
    b}) {\hat a}_{\rm b}^i {\hat
    a}_{\rm b}^j/2\right],
\label{hijdef}
\ee
with ${\hat a}_{\rm b}^i = {\bar a}_{\rm b}^i / {\bar a}_{\rm b} = \partial_i \Phi / |
{\bfnabla \Phi}|$.  The quantity $\rho_{T0}$ is the khronon mass
density (\ref{rhoTdef}) evaluated at $\pi = 0$:
\be
\rho_{T0} = \left. \rho_T \right|_{\pi=0}.
\label{rhoT0def}
\ee
We see that the khronon field is generated when there is a time
varying potential $\Phi$.

We can use the linearized equation (\ref{dynamics}) to make an
order of magnitude estimate of $\pi$ for self-gravitating sources.
In the MOND regime ${\bar a}_{\rm b} \ll a_0$ we obtain ${\bar \chi} \sim 1$,
$h^{ij} \sim G^{-1}$, $\nabla \Phi \sim \sqrt{G {\cal M} a_0} {\cal
  L}^{-1}$, $\rho_{T0} \sim \sqrt{{\cal M} a_0/G} {\cal L}^{-2}$ and
  ${\dot \rho}_{T0} \sim \rho_{T0} {\cal T}^{-1}$.  From footnote
\ref{calT} we find that the two terms on the left hand side of
Eq.\ (\ref{dynamics}) are comparable.  Comparing the second term to
the right hand side then gives the estimate
\be
\pi \sim {\cal L}^2 {\cal T}^{-1}.
\label{piest}
\ee
Inserting this estimate into the formula (\ref{Xidef}) for the potential
$\Xi$ we see that all three terms are comparable\footnote{In particular the term
nonlinear in $\pi$ in Eq.\ (\ref{Xidef}) is comparable to
the linear terms, which implies that the omitted nonlinear terms in
the linearized equation (\ref{dynamics}) should be included.  
However that linearized equation should be adequate for the order
of magnitude estimates derived here.}.
From Eq.\ (\ref{eom0})
it then follows that the khronon corrections to the MONDian dynamics
are of order unity, and that the acceleration of the khronon foliation
$\sim \nabla {\dot \pi}$ is of order the characteristic acceleration
of the system ${\cal L} {\cal T}^{-2}$.
This disproves the conjecture
by BM that the 
preferred foliation essentially coincides with the cosmological rest
frame with an acceleration
${\bfnabla} {\dot \pi}$ small compared to $a_0$ and
unimportant dynamically when ${\bar a}_{\rm b} \sim a_0$.

A similar analysis can be carried out in the high acceleration or
Newtonian regime ${\bar a}_{\rm b} \gg a_0$.  In this case
we have $| {\bar \chi} | \ll 1$, $\Phi \sim G {\cal M} {\cal L}^{-1}$,
$\rho_{T0} \sim {\bar \chi} {\cal M} {\cal L}^{-3}$ and $h^{ij} \sim
{\bar \chi} / G$.  The two terms on
the left hand side of Eq.\ (\ref{dynamics}) are again comparable and
we again find the estimate (\ref{piest}).  Although the perturbations
to the foliation are large in this regime, the effect of the khronon
field on the dynamics is suppressed by the small parameter ${\bar
  \chi}$ in Eq.\ (\ref{eom0}) which goes to zero as ${\bar a}_{\rm b} \to
\infty$ \footnote{Note that the validity of the approximation $\pi =0$
which we are discussing here is unrelated to the question of whether
one can work in the unitary gauge (\ref{Texpand1}), discussed in
Sec.\ \ref{sec:unitary} above.
In our analysis the gauge is determined by the conformally Cartesian
harmonic gauge condition with the expansion (\ref{metricPN}), and 
there is a physical, gauge-invariant quantity which reduces to $\pi$
in this gauge.}.

The special case of stationary solutions
with ${\dot \rho}_{\rm m} = {\dot \Phi} =0$
are consistent with a vanishing khronon field $\pi =0$, from
Eq.\ (\ref{dynamics}). 
We will study the stability of these solutions in
Sec.\ \ref{sec:stability} below.

\subsection{Consistency of the slow motion limit and subleading/post-Newtonian corrections}
\label{sec:subleading}

We next demonstrate that the assumed scalings (\ref{metricPN}) and
(\ref{Texpand}) that we have used for the slow motion limit give rise
to a consistent computational framework to subleading (post-Newtonian) order.  
Specializing to the gravitomagnetic sector, we find that the potential
$\zeta_i$ is given by (see the appendix)
\be
\nabla^2 \zeta_i = 16 \pi G ( \rho_{\rm m} v_{\rm m}^i + \rho_T v_T^i).
\label{sl1}
\ee
Here ${\bf v}_{\rm m}$ is the matter velocity and we have assumed for
simplicity that the matter is a fluid. We have also used the harmonic
gauge condition specified before Eq.\ (\ref{metricPN}), which implies that
$\partial_i \zeta_i = 4 {\dot \Phi}$.
Equation (\ref{sl1}) is the standard harmonic gauge equation of general relativity
but with 
the mass current
supplemented by a khronon contribution.
A similar calculation in the gravitoelectric sector shows that the
potential $\psi$ is given by (see the appendix)
\be
\partial_i \left\{ \left[ (1 + {\bar \chi}) \delta^{ij} + \left( \frac{1}{2}
        {\bar f}'' - {\bar\chi} \right) {\hat a}^i {\hat a}^j  \right]
\partial_j \psi \right\} = \ldots.
\label{sl2}
\ee
Here the ellipses represent source terms that are independent of $\psi$,
and ${\hat a}^i = {\bar a}^i/{\bar a}$ with ${\bar a}^i = \partial_i
\Xi$. The subleading equations (\ref{sl1}) and (\ref{sl2})
yield unique solutions for the potentials $\psi$ and $\zeta_i$
and place no constraints on the leading order fields.

\subsection{Comparison with other treatments of the slow motion limit in the literature}

\citet{Blanchet:2011wv} derived the slow motion limit of their theory
in the unitary gauge (\ref{Texpand1}).
However, they assumed the standard scaling (\ref{metricPN}) for $g_{ti}$, which as we
argued in Sec.\ \ref{sec:unitary} is inconsistent with the assumption of unitary gauge.
Nevertheless, this inconsistency did not affect their derivation of the
leading order MOND equation of motion.

A detailed analysis of the slow motion limit of khronometric
theories has been given by \citet{Bonetti:2015oda}.
They also assume unitary gauge and the standard post-Newtonian
scalings with $c$ of the 
metric coefficients, in disagreement with Eq.\ (\ref{gmexpand}).
However in their case the assumption is justified, because they work in
the context of a more general class of theories obtained by adding to
the action (\ref{action0}) the terms
\be
-\frac{c^3}{16 \pi G} \int d^4x \sqrt{-g} \left[ \beta \sigma_{ab}
  \sigma^{ab} + (\lambda + \beta/3) \theta^2 \right],
\ee
where $\beta$ and $\lambda$ are dimensionless parameters.  In the slow
motion limit this action reduces to 
\be
-\frac{c^2}{16 \pi G} \int d^4x \left[ \beta \pi_{,ij} \pi_{,ij}
  + (\lambda + \beta/3) (\bfnabla^2 \pi)^2 \right].
\label{BB1}
\ee
Comparing with the original action (\ref{action1}) we see that the
correction (\ref{BB1}) is superleading in the limit $c \to 0$,
scaling $\propto c^2$ rather than $c^0$, and
thus changing the nature of the post-Newtonian expansion.  The slow
motion limit $c \to \infty$ does not commute with the limit $\beta,
\lambda \to 0$ in which the theory reduces to BM theory.
The expansion used by \citet{Bonetti:2015oda} is valid in the limit
where the dimensionless parameters
\be
\beta {\hat c}^2, \ \ \ \lambda {\hat c}^2
\label{key}
\ee
are large compared to unity,
as they point out in their Sec.\ V, 
and the different expansion derived here
is valid in the limit when these parameters are small.
Here the parameter ${\hat c}$ is defined in Eq.\ (\ref{hatcdef}).

The expansion method of \citet{Bonetti:2015oda} gives results for some
of the post-Newtonian fields that diverge as $\beta, \lambda\to0$.
They describe this situation as a ``strong coupling'' and a breakdown
of the post-Newtonian expansion, a pathological limit.\footnote{One
example given by \citet{Bonetti:2015oda} that demonstrates the
breakdown of their expansion in the $\lambda, 
\beta \to 0$ limit is the following.
Their gravitomagnetic equation (44) multiplied by $\beta + \lambda$ and then
evaluated at $\beta = \lambda =0$ enforces the usual Poisson equation
relating the Newtonian potential and the matter mass density.  This
Poisson equation is inconsistent with the MOND equation for the
Newtonian potential, their Eq.\ (39).}  In fact the limit is well
defined but does require switching to the different kind of
post-Newtonian expansion used here (which is consistent at $\beta =
\lambda=0$) once the parameters (\ref{key}) become 
small.

\section{Stability of stationary solutions}
\label{sec:stability}

As discussed in Sec.\ \ref{sec:slomo}, stationary solutions\footnote{Here
by stationary solutions we mean solutions in which the gravitational
degrees of freedom $\pi$ and $\Phi$ are independent of time.
We do
not impose, for example, that fluid velocities should vanish, {\it i.e.} that
the solutions be static.}
of modified
Newtonian dynamics coincide with stationary solutions of the slow
motion limit of BM theory
with vanishing khronon perturbation \citep{Blanchet:2011wv}.
In this section we show that perturbations to these
solutions are stable in the
low acceleration regime ${\bar a} \ll a_0$, for the
cases of spherical, cylindrical and planar symmetry.  

Consider a stationary solution with $\pi = 0$.  Expanding the action
(\ref{action1}) to second order in the perturbations $\delta \pi$ and $\delta \Phi$
about this solution 
and making use of Eqs.\ (\ref{a2}), (\ref{Xidef}), (\ref{rhoTdef}) and (\ref{barchidef})
gives the quadratic action
\be
S_2[\delta \pi, \delta \Phi, \ldots] = S_{2,\pi}[\delta \pi] +
S_{2,{\rm int}}[\delta \pi, \delta \Phi] + S_{2,\Phi}[\delta \Phi,
  \ldots].
\label{actionpert}
\ee
Here the ellipses $\ldots$ represent the matter degrees of freedom, which in the application to cosmology would be the baryons.  The interaction term is
\begin{eqnarray}
  S_{2,{\rm int}} &=& \int dt \int d^3x \, {\dot {\delta \pi}} \delta \rho_T[\delta \Phi] \nonumber \\
 & =&
\int dt \int d^3x \, {\dot {\delta \pi}} \partial_i (h^{ij} \partial_j
\delta \Phi),
\label{interaction}
\end{eqnarray}
while the khronon action is
\be
S_{2,\pi} = \frac{1}{2} \int dt \int d^3 x \, \left[ h^{ij} \partial_i
  {\dot {\delta \pi}} \partial_j {\dot {\delta \pi}} - \rho_{T0}
  ({\bfnabla} \delta \pi)^2\right].
\label{action2}
\ee
Here the background khronon mass density $\rho_{T0}$ is given by
Eq.\ (\ref{rhoT0def}) and the tensor $h^{ij}$ is defined in
Eq.\ (\ref{hijdef}).  The tensor $h^{ij}$ is nonnegative if the
function ${\bar 
  f}({\bar a})$ obeys the conditions 
\be
   {\bar f}'({\bar a}) \le 0, \ \ \ \ {\bar f}''({\bar a}) \le 0,
   \label{condt1}
\ee
from Eq.\ (\ref{barchidef}).
These conditions are satisfied in the deep MOND regime ${\bar a} \ll a_0$ 
from Eqs.\ (\ref{fdef}) and (\ref{barfdef}).  Thus the khronon kinetic energy term in the action
(\ref{action2}) has the conventional sign in this regime.

We will assume that the background stationary solution is stable in the
conventional MOND theory without a khronon field \citep{1983ApJ...270..365M}.
This implies that 
the dynamics arising from the 
the term $S_{2,\Phi}$ in the action (\ref{actionpert}) has no unstable modes.

To show that that the theory (\ref{actionpert}) has no unstable modes
we proceed in two steps.  First, we show that the interaction term
$S_{2,{\rm int}}$ cannot produce an instability if the khronon term
$S_{2,\pi}$ is stable by itself.  Second, we analyze the dynamics of
the khronon field by itself and show that it has no unstable modes.

To analyze the interaction term $S_{2,{\rm int}}$ we rewrite the
action (\ref{actionpert}) in terms of the schematic Lagrangian
\be
L = {1 \over 2} {\cal G}_{AB} {\dot Q}^A {\dot Q}^B - {1 \over 2}
{\cal V}_{AB} Q^A Q^B + {\cal B}_{AB} Q^A {\dot Q}^B,
\label{ll}
\ee
where ${\cal B}_{AB} = {\cal B}_{[AB]}$.
Here $Q^A$ are configuration space coordinates that encompass $\delta \pi$, $\delta \Phi$
and the perturbations to the matter degrees of freedom, with $Q^A=0$
for the stationary solution.  The indices $A,B, \ldots$ label these fields and also
parameterize the dependence on the spatial coordinates, so sums over these indices
contain integrals over the spatial coordinates.
The third term in the Lagrangian (\ref{ll}) involving the
antisymmetric tensor ${\cal B}_{AB}$
contains
the interaction term
$S_{2,{\rm int}}$, which from Eq.\ (\ref{interaction}) is a product of two terms,
one of which has a time derivative.  See the appendix for more details
about the schematic Lagrangian (\ref{ll}).

The corresponding Hamiltonian can be written as
\be
H = \frac{1}{2} {\cal G}^{AB} ( P_A - {\cal B}_{AC} Q^C) ( P_B - {\cal
  B}_{BD} Q^D) + \frac{1}{2} {\cal V}_{AB} Q^A Q^B,
\ee
with ${\cal G}^{AB} {\cal G}_{BC} = \delta^A_C$. 
A key point now is that the 
tensors ${\cal G}_{AB}$ and ${\cal V}_{AB}$ are nonnegative.
For the non-khronon contributions to these tensors, this follows from our
assumption that the background stationary solution is stable in the
conventional MOND theory without a khronon field \citep{1983ApJ...270..365M}.
For the khronon contributions, the non-negativity of ${\cal G}_{AB}$
and ${\cal V}_{AB}$ follows from the discussion around
Eq.\ (\ref{condt1}) for the kinetic term, and from the form of the
second term in Eq.\ (\ref{action2}) for the potential term (assuming
$\rho_{T0} \ge 0$; see below).
It follows that $H$ is a non-negative quadratic form on phase space,
and so the motion is confined to a compact surface $H = $ constant.
This excludes any exponentially growing mode solutions of the form
$Q^A(t) = Q^A_0 e^{-i \omega t}$.

Turn now to the dynamics of the khronon field by itself, described by
the action (\ref{action2}).
Consider complex mode solutions of the form
\be
\delta \pi(t,{\bf x}) = e^{-i \Omega t} \pi_0({\bf x}).
\ee
Substituting this ansatz into the equation of motion obtained from the
action (\ref{action2}), multiplying by $\pi_0({\bf x})^*$ and
integrating by parts yields
\begin{eqnarray}
\Omega^2 = \frac{ \int d^3 x\, \rho_{T0} | {\bfnabla } \pi_0 |^2}{\int d^3
  x \, h^{ij} \partial_i \pi_0 \partial_j \pi_0^* }.
\label{pp}
\end{eqnarray}
It follows that the background MOND solution has no unstable modes if
(i) the conditions (\ref{condt1}) are satisfied, so that $h^{ij}$ is
nonnegative, and (ii) 
the background khronon mass density given by Eqs.\ (\ref{rhoTdef}) and
(\ref{rhoT0def}) is nonnegative:
\be
\rho_{T0} \ge 0.
\label{condt2}
\ee

The khronon mass density is not always nonnegative for stationary
solutions, as shown by \citet{1986ApJ...306....9M}.
In particular he showed that if there exists
an isolated point where ${\bfnabla} \Phi=0$, located in a region with
$\rho_{\rm m}
=0$, then that point must lie on the boundary of a region with with $\rho_{T0} < 0$.
We now show that $\rho_{T0} \ge 0$ when $\pi=0$ in the cases of spherical, cylindrical and
planar symmetry, thus showing that the background solutions are mode stable
in the deep MOND regime in these cases.  Stability going beyond these special cases
is an open question\footnote{Note that the fact that $\rho_{T0}$ is
sometimes negative does not imply the existence of an instability,
since the contributions to the integral in the numerator of Eq.\ (\ref{pp}) from regions of
positive $\rho_{T0}$ may dominate.}.

In spherical symmetry the magnitude $g(r)$ of the Newtonian
acceleration at radius $r$ is related to the enclosed mass $m(r)$ by
$g = G m(r)/r^2$.  The magnitude ${\bar a}$ of the actual acceleration
is given in terms of $g$ from Eq.\ (\ref{eom1}) by $\varpi({\bar a}) = g$,
where the function $\varpi$ is given by\footnote{The conventional notation for
this function is $\varpi({\bar a}) = {\bar a} \mu( {\bar a}/a_0)$
\citep{1983ApJ...270..365M}.}
\be
\varpi({\bar a}) = {\bar a} + {\bar a} {\bar \chi}({\bar a}).
\label{varpidef}
\ee
We write this relation as ${\bar a} =
\lambda(g)$, where $\lambda$ is the inverse of the function $\varpi$.
Using the fact that the total effective mass ${\tilde m}(r)$ enclosed
inside radius $r$ is given by ${\bar a} = G {\tilde m}(r)/r^2$,
we obtain
\be
\frac{G {\tilde m}(r)}{r^2} = \lambda \left[ \frac{G { m}(r)}{r^2}
  \right].
\label{relation}
\ee
Now the khronon mass density $\rho_{T0}$ is proportional to ${\tilde
  m}'(r) - m'(r)$, which from Eq.\ (\ref{relation}) is given by
\be
   {\tilde m}'(r) - m'(r) = \frac{2 r}{G} \left[ \lambda(g) - g
     \lambda'(g) \right] + m'(r) \left[ \lambda'(g) -1 \right].
\ee
It follows that $\rho_{T0} \ge 0$ if the conditions
\be
\lambda'(g) \ge 1,  \ \ \ \ \lambda(g) - g \lambda'(g) \ge 0,
\ee
are satisfied.  From Eqs.\ (\ref{barchidef}) and (\ref{varpidef}) these conditions are
equivalent to
\be
{\bar f}''({\bar a}) \le 0, \ \ \ \ \ {\bar a} {\bar f}''({\bar a}) -
{\bar f}'({\bar a}) \ge 0.
\label{final}
\ee
Both of these conditions are satisfied in the deep MOND regime ${\bar a} \ll a_0$ 
from Eqs.\ (\ref{fdef}) and (\ref{barfdef}).

Similar analyses apply in the cases of cylindrical and planar
symmetry.  In the cylindrical case the relation (\ref{relation}) is modified
by replacing on both sides $G m(r)/r^2$ with $2 G \sigma(r)/r$, where
$r$ is now distance from the axis of symmetry, and $\sigma(r)$ is mass
per unit length along the axis enclosed inside radius $r$.
Repeating the analysis gives the same conditions (\ref{final}) as
before.  For planar symmetry the argument of the function $\lambda$
becomes $4 \pi G \Sigma(r)$, where $r$ is now distance from the plane
and $\Sigma(r)$ is enclosed mass per unit area.  In this case one
obtains the first of the conditions (\ref{final}) but not the second.

To summarize, we have shown that 
that stationary solutions are stable if:
\begin{enumerate}
\item We restrict to special configurations of enhanced symmetry.
\item The solutions are stable in the conventional MOND theory without
  the khronon field.
 \item The conditions (\ref{condt1}) and (\ref{final}) on the function ${\bar f}$ are
satisfied.  
\end{enumerate}

We can combine the two conditions on ${\bar f}$ and write them in terms of
the function $f$ that appears in the action (\ref{action0}) using the
rescaling (\ref{barfdef}), to obtain
\vspace{-0.1cm}
\begin{eqnarray}
f'(a) \le a f''(a) \le 0.
\label{final2} 
\end{eqnarray}
Although these conditions are satisfied in the deep MOND regime, the
cannot be satisfied for all values of $a$ since they are incompatible
with the boundary conditions (\ref{fdef}) and (\ref{fdef1}) at large $a$ and
small $a$.  Thus there must be a range of values of $a$ with $a \sim
a_0/c^2$ or ${\bar a} \sim a_0$ where the conditions (\ref{final2}) are violated.
Stationary solutions might therefore be unstable in this regime\footnote{If instabilities do exist, the growth timescale when ${\bar a} \sim
a_0$ will be of order the gravitational dynamical timescale ${\cal
  T}$, as can be seen by applying dimensional analysis to the action
(\ref{action2}) (see footnote \ref{calT} above).}.

\section{Discussion and Conclusions}\label{sec:conclusions}

The BM theory of modified Newtonian dynamics \citep{Blanchet:2011wv,Blanchet:2012ub}
is a minimal fully relativistic version of MOND which is simpler than
the alternatives \citep{Skordis:2020eui,Skordis:2021mry}.
In this paper we have presented arguments in favor of the viability of
this theory.

In particular, it was previously shown by BM that stationary solutions of this theory in the
slow motion limit coincide with stationary solutions of the
conventional nonrelativistic formulation of MOND.  The theory would be
disfavored if these solutions were unstable to khronon perturbations.
We have presented evidence for stability of these solutions in certain limits.

We also showed that predictions of the BM theory generally differ from
those of MOND by an amount of order unity for non-stationary
solutions, for systems near the MOND acceleration scale.  In
particular this applies to binary star systems, for which the khronon
perturbation is nonzero and is determined by 
an elliptic equation in the rotating frame. It would be interesting
to derive from this equation the form of Kepler's third law for the BM
theory, which will differ from the MOND form \citep{Zhao:2010yu}.
This could be useful 
for current efforts to test MOND with observations of wide binary star
systems using GAIA data \citep{Pittordis:2022qrz,Chae:2023prf}.

We also note that instabilities arise in the BM theory when perturbed
about Minkowski spacetime. These instabilities are generic for all khronometric
theories and can be cured by the addition of higher spatial derivative
terms to the action which are suppressed by a mass scale, and give
small corrections the dynamics at scales of interest 
\citep{Horava:2009uw,Blas:2009qj,Bonetti:2015oda}.

\begin{acknowledgements}

I thank Enrico Barausse helpful correspondence and for pointing out an
error in an earlier version of this paper, Ira Wasserman for
helpful discussions, and an anonymous referee for detailed and helpful comments.
This research was supported in part by NSF grant
PHY 2110463 and by a fellowship from the Simons Foundation.

\end{acknowledgements}

\begin{appendix}

\setcounter{equation}{54}
  
In this appendix we provide some of the details of the calculations
reported in the body of the paper.

We obtain the expansion (\ref{n1}) of the normal $n_a$ by inserting the ansatz
(\ref{Texpand}) for the khronon field $T$ into the definition (\ref{ndef}) and using the metric expansion (\ref{metricPN}). 
Inserting the result into the definition (\ref{adef}) of the acceleration and using the expansion (\ref{metricPN}) again gives
\be
   {\vec a} = \frac{1}{c^2} \partial_i \Xi \frac{\partial}{\partial
     x^i} + O \left( \frac{1}{c^4} \right),
   \label{aa2}
   \ee
where $\Xi$ is given by Eq.\ (\ref{Xidef}).  Using the rescaling
(\ref{rescale}) now gives the formula (\ref{a2}).  

Next, we insert the acceleration (\ref{aa2}) and normal (\ref{n1})
into the khronon stress energy expression (\ref{khrononTab}), and make use of
the rescalings (\ref{barfdef}) and (\ref{barchidef}).  This shows that
the stress energy components have an expansion with $c$ of the
standard post-Newtonian form (\ref{standard}), with the leading order
terms being
  \begin{eqnarray}    
  \label{khrononst}
    T_T^{tt} &=& \rho_T + O\left( \frac{1}{c^2} \right), \\
    \label{mf}
    T_T^{ti} &=& \rho_T v_T^i + O\left( \frac{1}{c^2} \right), \\
    \label{Tij}
    T_T^{ij} &=& \frac{1}{8 \pi G} \left[ - {\bar f} \delta_{ij} + 2
      {\bar \chi} \Xi_{,i} \Xi_{,j} \right] - \rho_T \pi_{,i} \pi_{,j}  + O\left( \frac{1}{c^2} \right), 
  \end{eqnarray}
 where $\rho_T$ is the khronon mass density (\ref{rhoTdef}) and ${\bf v}_T$
 its velocity (\ref{vTdef}).

We next turn to the subleading corrections to the slow motion limit
discussed in Sec.\ \ref{sec:subleading}.
Substituting the momentum flux (\ref{mf}) together with the fluid
momentum flux into the usual
post-1-Newtonian harmonic gauge gravitomagnetic Einstein equation
gives Eq.\ (\ref{sl1}).  In the gravitoelectric sector, the subleading
equation for the scalar potential $\psi$ in general relativity in
conformally Cartesian harmonic gauge is
\be
c^6 R^{tt} = \nabla^2 \psi - {\ddot \Phi} = 4 \pi G [T^{(2)\,tt} +
\delta_{ij} T^{(0)\,ij}].
\ee
Here on the right hand side $T^{(0)\,\alpha\beta}$ is the $O(c^0)$
piece of the stress energy tensor, and $T^{(2)\,\alpha\beta}$ is the
$O(c^{-2})$ piece.  We can apply this equation to the present context
by inserting on the right hand side the sum of the fluid stress energy
tensor and the khronon stress energy tensor (\ref{khrononst}) -- (\ref{Tij}).
Bringing all the terms that depend on $\psi$ to the left hand side
then results in Eq.\ (\ref{sl2}).
Because of how $\psi$ enters into the metric expansion
(\ref{metricPN}), the relevant terms can be computed by taking a
variation $\Phi \to \Phi + \delta \Phi$ of the expression (\ref{rhoTdef}).
Note that the right hand side of
Eq.\ (\ref{sl2}) will then depend on subleading corrections to the
khronon field expansion (\ref{Texpand}) beyond the field $\pi$,
arising from $T_T^{(2)\,tt}$.

Finally we provide more details about the schematic Lagrangian
(\ref{ll}) that describes perturbations to stationary solutions.
The kinetic energy term (first term) in the Lagrangian (\ref{ll}) has
a contribution from the khronon field given by the first term in
Eq.\ (\ref{action2}).  It also has a contribution from the matter
degrees of freedom, the third term in Eq.\ (\ref{actionpert}).
The specific form of this term for a fluid, for example, in the
framework of Lagrangian perturbation theory, is given in Eq.\ (6.6.7)
of \citet{shapiro2008black}.
The potential term (second term) in the Lagrangian (\ref{ll}) has the khronon
contribution given by the second term in Eq.\ (\ref{action2}), and the
fluid contribution given by Eq.\ (6.6.9) of
\citet{shapiro2008black}.  To get the correct dependence on $\delta
\Phi$ for our context we multiply the last term in this equation by
two\footnote{This is necessary since $\delta \Phi$ has effectively been
integrated out in \citet{shapiro2008black}, and we are restoring it as
a dynamical variable.}, and add the
contribution
\be
\frac{1}{2} \int dt \int d^3 x h^{ij} \partial_i \delta \Phi
\partial_j \delta \Phi.
\ee
Finally the mixed term (third term) in
(\ref{ll}) has the khronon contribution (\ref{interaction}).
It can also have a fluid contribution, for perturbations to stationary
systems that are not static \citep{1967MNRAS.136..293L}.

\end{appendix}

\bibliographystyle{aasjournal}
\bibliography{Ref1}

\end{document}